\documentclass{jpp}
\usepackage{subeqn}
\usepackage{epsfig}

\let\realverbatim\verbatim
\let\realendverbatim\endverbatim
\renewenvironment{verbatim}
  {\par\addvspace{6pt plus 1pt minus 2pt}\realverbatim}
  {\realendverbatim\addvspace{6pt plus 1pt minus 2pt}}

\newcommand\dynpercm{\nobreak\mbox{$\;$dynes\,cm$^{-1}$}}

\newcommand\Real{\mbox{Re}}     % cf plain TeX's \Re and Reynolds number
\newcommand\Imag{\mbox{Im}}     % cf plain TeX's \Im
\newcommand\Rey{\mbox{\textit{Re}}}  % Reynold's number
\newcommand\Pran{\mbox{\textit{Pr}}} % Prandtl's number, cf plain TeX's \Pr product
\newcommand\Ai{\mbox{\textit{Ai}}}
\newcommand\Bi{\mbox{\textit{Bi}}}

\ifprodtf \else
  \checkfont{eurm10}
  \iffontfound
    \IfFileExists{upmath.sty}
      {\typeout{^^JFound AMS Euler Roman fonts on the system,
                   using the 'upmath' package.^^J}%
       \usepackage{upmath}}
      {\typeout{^^JFound AMS Euler Roman fonts on the system, but you
                   don't seem to have the}%
       \typeout{'upmath' package installed. JPP.cls can take advantage
                 of these fonts,^^Jif you use 'upmath' package.^^J}%
       \providecommand\upi{\pi}%
       \providecommand\umu{\umu}%
      }
  \else
    \providecommand\upi{\pi}%
    \providecommand\umu{\mu}%
  \fi
\fi

\ifprodtf \else
  \checkfont{msam10}
  \iffontfound
    \IfFileExists{amssymb.sty}
      {\typeout{^^JFound AMS Symbol fonts on the system, using the
                'amssymb' package.^^J}%
       \usepackage{amssymb}%

      }{}
  \else
  \fi
\fi

\ifprodtf \else
  \IfFileExists{amsbsy.sty}
    {\typeout{^^JFound the 'amsbsy' package on the system, using it.^^J}%
     \usepackage{amsbsy}}
    {\providecommand\boldsymbol[1]{\mbox{\boldmath $##1$}}}
\fi

\newcommand\ssC{\mathsf{C}}    % for sans serif C
\newcommand\sfsP{\mathsfi{P}}  % for sans serif sloping P
\newcommand\bmath[1]{\ensuremath{\boldsymbol{#1}}}

\newcommand{\be}{\begin{equation}}
\newcommand{\ee}{\end{equation}}

\newcommand{\eqa}{\begin{eqnarray*}}
\newcommand{\eqe}{\end{eqnarray*}}
\newcommand{\eqnu}{\begin{eqnarray}}
\newcommand{\eqne}{\end{eqnarray}}

\newcommand{\eeq}{\end{eqnarray}}

\newdefinition{definition}[theorem]{Definition}

\title[Journal of Plasma Physics]
{Self-Organization  in a  Driven Dissipative Plasma System}

\author[Shaikh et al]
{D\ls A\ls S\ls T\ls G\ls E\ls E\ls R \ns S\ls H\ls A\ls I\ls K\ls$^{1,2}$
H\ls , B.\ls D\ls A\ls S\ls G\ls U\ls P\ls T\ls A\ls$^{2}$,   Q.\ls H\ls U\ls$^{2}$ and G.\ls
P.\ls Z\ls A\ls N\ls K\ls$^{1,2}$}
\affiliation{$^{1}$Department of Physics and \\
$^{2}$Center for Space Physics and Aeronomic Research (CSPAR),\\
University of Alabama at Huntsville, Huntsville, AL 35805. USA.\\
{\tt Email:dastgeer.shaikh@uah.edu}}
\date{12 March 2009, Revised 25 March 2009, Accepted 25 March 2009}
\pubyear{2009} % only needed when year of publication is not current year.
\volume{00}
\part{0}
\pagerange{\pageref{firstpage}--\pageref{lastpage}}
\doi{S0963548301004989}

\begin{document}

\label{firstpage}
\maketitle

\begin{abstract}

We perform a fully self-consistent 3-D numerical simulation for a
compressible, dissipative magneto-plasma driven by large-scale
perturbations, that contain a fairly broader spectrum of
characteristic modes, ranging from largest scales to intermediate
scales and down to the smallest scales, where the energy of the
system are dissipated by collisional (Ohmic) and viscous
dissipations. Additionally, our simulation  includes nonlinear
interactions amongst a wide range of fluctuations that are
initialized with random spectral amplitudes,  leading to the
cascade of spectral energy in the inertial range spectrum, and
takes into account large scale as well as small scale perturbation
that may have been induced by the background plasma fluctuations,
also the non adiabatic exchange of energy leading to the migration
of energy from the energy containing modes or randomly injected
energy driven by perturbations and further dissipated by the
smaller scales. Besides demonstrating the comparative decays of
total energy and dissipation rate of energy, our results show the
existence of a perpendicular component of current, thus clearly
confirming that the self-organized state is non-force free.

\end{abstract}

%\tableofcontents

%\begin{center}
\section{Introduction and Motivation:}
%\end{center}

The phenomena of self-organization, in which a continuous system
naturally evolves towards a state exhibiting some form of order on
large scales, is deeply rooted in nature.  The ordered states are
remarkably robust; their detailed structure remains relatively
invariant across experimental realization; these preferred states
are independent of the way the system is prepared (Hasegawa, 1985;
Ortolani and Schnack, 1993)

Several broad classes of physical processes , such as first and second
order phase transitions, crystallization, structure formations in
space and astrophysics and cosmology can be described as examples of
self-organization.  In chemistry, examples of self-organizations
include reaction-diffusion system such as Belousov-Zhabotinsky
reactions, self-assembled monolayers, molecular self-assembly,
etc. Biological systems are swarmed with examples of
self-organizations; besides the example of pattern formation and
morphogenesis, the origin of life itself from self-organizing chemical
systems is the supreme example self-organization. Most of the system
exhibiting self-organization in nature are intrinsically nonlinear and
often are not isolated - they are driven by some external input.
Self-organization occurs in open systems which are far from thermal
equilibrium. Nicolis and Prigogine (1977) had envisaged new types of
self-organized states for driven dissipative systems and called these
states as "Dissipative Structures".  Such structures provide striking
examples of non-equilibrium as a source of order.  During the past
decades, non-linear dynamics and non- equilibrium thermodynamics have
developed along with plasma physics, providing qualitatively new
approaches to complex problems. It is thus widely accepted nowadays,
for instance, that non-equilibrium may be a source of order in
dissipative systems, allowing the emergence of self-organization
Self-organizations have been observed in varieties of laboratory
magnetically confined systems.  Most plasma systems are dissipative,
externally driven and far from equilibrium, thus sharing some
essential features with other complex non-linear systems that show
spatial and temporal coherence.  Such a system is described by a set
of non-linear partial differential equations.  During its relaxation
towards a self-organized state, it takes advantage of instabilities
that lead it, for instance, to a preferred state, under certain
constraints that prevent it from falling into a trivial unconfined
state. Such constraints appear as quadratic or higher order
quantities, which are conserved in the absence of dissipation. The
relevant feature is that, in the presence of dissipation one of these
quantities decays faster than the others. Generally, it remains to be
determined which is the relevant variational principle underlying the
relaxation mechanisms described above, is appropriate to characterize
the self-organized state. In the space and solar plasma flows, a
number of commonly observed processes including flares, prominences,
filaments and/or (magnetic) loop like structures that are generated
often during the active solar period can be modelled by relaxation of
plasma through self organization (Bhattacharya et al., 2007).  The
kinematic as well as dynamics of these entities are far more complex
than ever thought and we still lack an in depth insight into their
evolutionary characteristics despite the availability of an enormous
data bases from various spacecraft missions. A part of the problem
lies with the lack of a fully self-consistent description (or physical
model) of evolution of these structures in realistic environments. For
instance, one of the mechanisms of the generation of magnetic field
loops is often attributed to the Taylor relaxation (Taylor, 1974)
process where magnetic stresses overcome the pressure stresses which
thereby balance all the MHD forces. Such state is often characterized
by a low plasma beta (where plasma-beta is a ratio of pressure and
magnetic energy).  Under no external forces, the pressure gradients
flatten out to nullify the magnetic pinch $({\bf J} \times {\bf B})$
force. This state is called a "Force Free State". The force free state
is one of the analytic descriptions that describes magnetic loop in
terms of a magnetic field configuration where rotation of magnetic
field is proportional to the field itself thus giving rise to a
constant of proportionality. The constant of proportionality can be a
real constant (linear force free) or dependent on space (nonlinear
force free). The validity of a force free model is however restricted
to the formation of the non evolutionary magnetic loops essentially in
the vicinity of the solar corona. Further they are strictly invalid to
describe the evolution dynamics. This further necessitates the
development of a self-consistent description of coronal magnetic field
loops that can respond to as well as interact with the realistic
perturbations ubiquitously present in the solar atmosphere. Motivated
by the issues described above, we have developed a more generic model
based on three dimensional simulations of fully compressible, non
adiabatic, driven-dissipative magnetofluid plasma that contains a
fairly broader spectrum of characteristic modes. The underlying modes
range from largest scale (of the size of the system) to the
intermediate scales (constituting the inertial range spectra) to the
smallest scales where energy of the system can be dissipated by virtue
of collisional or viscous dissipation. One of the novel features of
our 3D compressible MHD plasma model is that it deals with the entire
spectrum of fluctuations unlike those work that either describe single
mode of flux of coronal magnetic field and ignore the background small
scale realistic perturbations or single coherent structures (Amari and
Luciani, 2000). Additional features of our simulation model are that
it (i) includes nonlinear interactions amongst a wide range of
fluctuations that are initialized with random spectral amplitudes. The
nonlinearities in the underlying system drive turbulent processes and
lead to the cascade of spectral energy in the inertial range spectrum,
(ii) large scale as well as small scale perturbation that may have
been induced by the background plasma fluctuations, (iii) non
adiabatic exchange of energy leading to the migration of energy from
the energy containing modes or randomly injected energy driven by
perturbations and further dissipated by the smaller scales.
\\

The plan of the paper is as follows: In section 2 we briefly
summarize some of the MHD relaxation models appropriate for driven
dissipative plasma based on the principle of minimum dissipation
rate (MDR) leading to non-force free states. Section 3 presents a
description of our model starting with the conservative forms of
the MHD equations with applicable normalization procedures.
Section 4 describes our simulation results. Section 5 include
summary of our work, conclusion and discussions of future works,
%\begin{center}
\section{Non-force free Self-organized States from MHD Relaxation Models}
%\end{center}

The seminal work on plasma relaxation proposed by Taylor predicts
a force free state for a magnetized plasma.  For such states,  the
current {\bf J } is always along the direction of the magnetic
field {\bf B } and the perpendicular component of the current
$J_\perp = 0$.  So, from the force-balance equation ${\bf J}
\times {\bf B} = \nabla p$, it is seen that for force-free states,
the pressure gradient is zero, and such states are not suitable
for devices confining plasma by magnetic fields.

A small amount of resistivity, ingrained in any realistic plasma,
is essential to allow reconnective processes leading to
relaxation. In fact, dissipation, along with nonlinearity, is
universal in systems evolving towards self-organized states and it
is natural to assume that dissipation plays a decisive role in the
self-organization of a system.  Alternative models plasma
relaxation, based on the  principle of minimum dissipation rate
(MDR) of energy have been  proposed by several authors (Montgomery
and   Phillips,  1988;  Farengo,  and Sobehart, 1994, 1995;
Farengo and Kaputi, 2002;  Dasgupta et al., 1998, 2002, 2009;
Bhattacharya and Janaki, 2004).   The principle of minimum
dissipation rate is closely akin to the principle of minimum
entropy production rate of irreversible thermodynamics as
formulated  by Prigogine (Prigogine, 1946) . Relaxation of a
driven dissipative plasma has been formulated by Bhattacharya and
Janaki (2004).  As shown by these authors, the relaxed states
obtained from MDR are non-force free,  i.e., for these states,
$\nabla p \neq  0$.  A critical signature of such non-force free
state is that the perpendicular component of the current $J_\perp
\neq 0$. It may be mentioned that relaxation of a MHD plasma to a
non-force free state has been numerically demonstrated by Zhu et
al., (1995). One of the main objectives of this work is to
investigate the existence of non-zero  $J_\perp $ in a
self-organized states of a driven system.

\section{Description of the Simulation Model}
The fluid model describing nonlinear turbulent processes in the
magnetofluid plasma, in the presence of a background magnetic field,
can be cast into plasma density ($\rho_p$), velocity (${\bf U}_p$),
magnetic field (${\bf B}$), pressure ($P_p$) components according to
the conservative form
\be
\label{mhd}
 \frac{\partial {\bf F}_p}{\partial t} + \nabla \cdot {\bf Q}_p={\cal Q},
\ee
where,
\[{\bf F}_p=
\left[
\begin{array}{c}
\rho_p  \\
\rho_p {\bf U}_p  \\
{\bf B} \\
e_p
  \end{array}
\right],
{\bf Q}_p=
\left[
\begin{array}{c}
\rho_p {\bf U}_p  \\
\rho_p {\bf U}_p {\bf U}_p+ \frac{P_p}{\gamma-1}\bar{\bf I}+\frac{B^2}{2}\bar{\bf I}-{\bf B}{\bf B}  \\
{\bf U}_p{\bf B} -{\bf B}{\bf U}_p\\
e_p{\bf U}_p
-{\bf B}({\bf U}_p \cdot {\bf B})
  \end{array}
\right], \]
\[{\cal Q}=
\left[
\begin{array}{c}
0  \\
{\bf f}_M({\bf r},t) +\mu \nabla^2 {\bf U}_p+\eta \nabla (\nabla\cdot {\bf U}_p)  \\
\eta \nabla^2 {\bf B}  \\
0
  \end{array}
\right]
\]
and
\[ e_p=\frac{1}{2}\rho_p U_p^2 + \frac{P_p}{\gamma-1}+\frac{B^2}{8\pi}.\]
Equations (\ref{mhd}) are normalized by typical length $\ell_0$
and time $t_0 = \ell_0/V_A$ scales in our simulations such that
$\bar{\nabla}=\ell_0{\nabla}, \partial/\partial
\bar{t}=t_0\partial/\partial t, \bar{\bf U}_p={\bf
U}_p/V_A,\bar{\bf
  B} ={\bf B}/V_A(4\pi \rho_0)^{1/2}, \bar{P}=P/\rho_0V_A^2,
\bar{e}_p=e_p/\rho_0V_A^2, \bar{\rho}=\rho/\rho_0$. The bars are
removed from the normalized equations (1). $V_A=B_0/(4\pi \rho_0)^{1/2}$
is the Alfv\'en speed, and $\bar{\bf I}$ is a unit tensor.

The rhs in the momentum equation denotes a forcing functions (${\bf
  f}_M({\bf r},t)$) that essentially influences the plasma momentum at
  the larger length scale in our simulation model.  With the help of
  this function, we drive energy in the large scale eddies to sustain
  the magnetized turbulent interactions. In the absence of forcing,
  the turbulence continues to decay freely.  While the driving term
  modifies the momentum of plasma, we conserve density (since we
  neglect photoionization and recombination). The large-scale random
  driving of turbulence can correspond to external forces or
  instabilities for example fast and slow streams, merged interaction
  region etc in the solar wind, supernova explosions, stellar winds in
  the ISM, etc. The magnetic field evolution is governed by the usual
  induction equation and obeys the frozen-in-field theorem unless
  nonlinear dissipative mechanism introduces small-scale damping.
  Note carefully that MHD plasma momentrum equation contains nonlinear
  terms on the rhs. This means that mode coupling processes can
  potentially be mediated by nonlinear interactions, in addition to
  the damping associated with the small-scale turbulent motion. Thus
  nonlinear turbulent cascades are not only responsible for the
  spectral transfer of energy in the inertial range, but are also
  likely to damp the plasma motion in a complex manner. Nonetheless,
  the spatio-temporal scales in the nonlinear damping can be {\it
  distinct} from that of the linear dissipation.  We restrict forcing
  of plasma momentum fluctuations in a region specified by the
  wavenumbers $k<k_f$ such that energy is injected in the large scale
  plasma fluctuations. The driving is random in time and space.  The
  amplitude of driving force is also chosen randomly between 0 and
  1. We further make sure that the random number generator used for
  the driving force is a isotropic and uniform in the spectral space
  and does not lead to spectral anisotropy.

Turbulent-relaxation evolution studies in three (3D) dimensions
are performed to investigate the nonlinear mode coupling
interaction of a decaying compressible MHD turbulence described by
the closed set of Eqs (\ref{mhd}). For this purpose, we have
developed a full three dimensional (3D) compressible
magnetohydrodynamic (MHD) code (Shaikh et al., 2008). All the
fluctuations are initialized isotropically (no mean fields are
assumed) with random phases and amplitudes in Fourier space and
evolved further by integration of Eqs (\ref{mhd}) using a fully
de-aliased pseudospectral numerical scheme.  Fourier spectral
methods are remarkably successful in describing turbulent flows in
a variety of plasma and hydrodynamic (i.e non magnetized) fluids.
Not only do they provide an accurate representation of the fluid
fluctuations in the Fourier space, but they are also
non-dissipative. Because of the latter, nonlinear mode coupling
interactions preserve ideal rugged invariants of fluid flows,
unlike finite difference or finite volume methods. The
conservation of the ideal invariants (energy, enstrophy, magnetic
potential, helicity etc.) in turbulence is an extremely important
feature in general, and {\it particularly} in our simulations,
because these quantities describe the cascade of energy in the
inertial regime, where turbulence is, in principle, free from
large-scale forcing as well as small scale dissipation.  The
precise measurement of the decay rates associated with the MHD
invariants is therefore one the major concerns in the study of
MDR.  Dissipation is nonetheless added physically in our
simulations to push the spectral cascades further down to the
smallest scales and also to allow minimal dissipation.  The
evolution variables are discretized in Fourier space and we use
periodic boundary conditions.  The initial isotropic turbulent
spectrum was chosen to be close to $k^{-2}$ with random phases in
all three directions. The choice of such (or even a flatter than
-2) spectrum do not influence the dynamical evolution of the
turbulent fluctuations as final state in all our simulations leads
to the identical results that are consistent with the proposed
analytic theory.  The equations are advanced in time using a
second-order predictor-corrector scheme. The code is made stable
by a proper de-aliasing of spurious Fourier modes and choosing a
relatively small time step in the simulations. Additionally, the
code preserves the $\nabla \cdot {\bf B} =0$ condition at each
time step.  Our code is massively parallelized using Message
Passing Interface (MPI) libraries to facilitate higher resolution
in a 3D volume.  Kinetic and magnetic energies are also
equi-partitioned between the initial velocity and the magnetic
fields. The latter helps treat the transverse or shear Alfv\'en
and the fast/slow magnetosonic waves on an equal footing, at
least during the early phase of the simulations.\\

\section{Simulation Results:}

Our major focus in the paper is to study turbulent relaxation of
magnetofluid plasma through nonlinear interactions. For this
purpose, we let magnetized fully compressible MHD turbulence
evolve under the action of nonlinear interactions in which random
initial fluctuations, containing sources of free energy, lead to
the excitation of unstable modes. These modes often deviate the
initial evolutionary system substantially away from its
equilibrium state.  When the instability tends to saturate,
unstable modes lead to fully developed turbulence in which larger
eddies transfer their energy to smaller ones through a forward
cascade until the process is terminated by the small-scale
dissipation. During this process, MHD turbulent fluctuations are
dissipated gradually due to the finite Reynolds number, thereby
damping small scale motion as well.  The energy in the smaller
Fourier modes migrates towards the higher Fourier modes following
essentially the vector triad interactions ${\bf k} + {\bf p} =
{\bf q}$. These interactions involve the neighboring Fourier
components (${\bf k},{\bf p},{\bf q}$) that are excited in the
local inertial range turbulence. We conjecture in our earlier work
(Shaikh et al., 2008) that MDR plasma relaxes towards nonlinear
non force free state state through nonlinear evolution.  In our
simulations, we undertake this point in the presence of driven
turbulence.  Our simulations are carried out in the presence of a
mean magnetic field (taken along the $z$ direction).  One of the
ways we ensure the nonlinear non force free evolution is by
determining whether or not there develops any perpendicular
component of plasma currents. In the following, we explain why it
is essential to self-consistently produce a perpendicular
component of plasma current ($J_\perp$) that facilitates the
nonlinear non force free interactions.

The nonlinear interactions in magnetoplasma turbulence are determined
predominantly by $J \times B$ forces in plasma momentum
equation. Similarly, since $ V \propto B$ through ($J$), the same
nonlinear interactions also govern magnetic induction equation.  In
the presence of a mean magnetic field along the $z$ direction, the
parallel and perpendicular components can be denoted respectively by
$J_\parallel$ and $J_\perp$. For obvious reasons, $J_\parallel \times
B =0$ for magnetic field fluctuations that lie strictly along the the
$J_\parallel$ component of plasma current. Hence this term contributes
negligibly in the nonlinear interactions. By contrast, it is only the
$J_\perp$ component, emerging from the $J_\perp \times B$ term, that
contributes largely to the nonlinear interactions. Thus current
fluctuations orthogonal to the mean or fluctuating magnetic field
component predominantly govern the nonlinear interactions.

\begin{figure}
\includegraphics{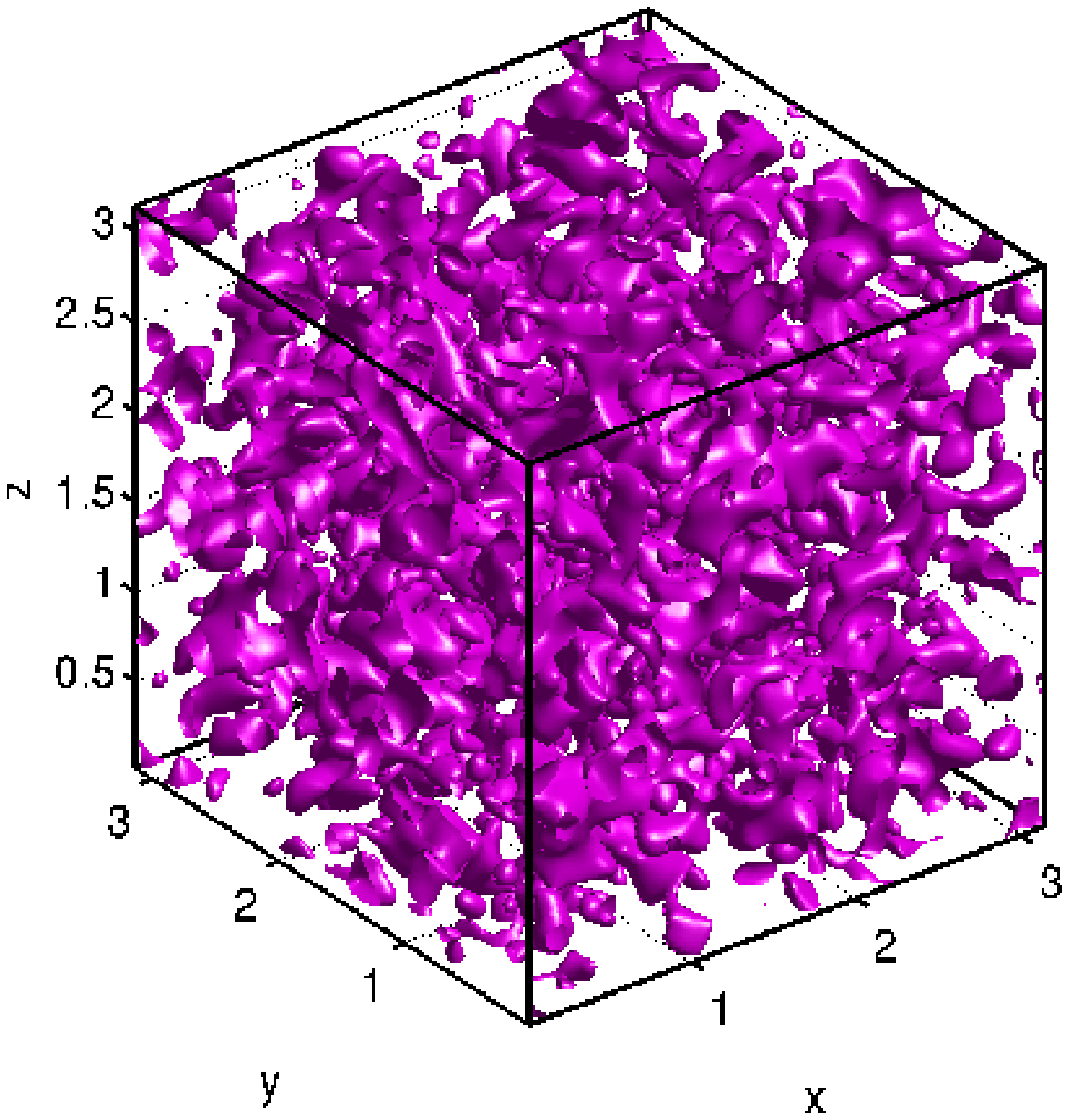}% Here is how to importEPS art
\caption{\label{fig1}Evolution of random initial turbulent
fluctuations lead to the formation of relatively small-scale
isotropic structures in 3D compressible MHD simulations. The
numerical resolution is $128^3$ in a cubic box of volume $\pi^3$.
The dissipation parameter $\eta = \nu = 10^{-4}$. Shown are the
iso-surfaces of the $x$-component of the magnetic field.}
\end{figure}

 Figure \ref{fig1} shows iso-surfaces of the $x$-component of the
 magnetic field by the evolution of random initial turbulent
 fluctuations leading to the formation of relatively small-scale
 isotropic structures.  Fig 1 is a typical snap shot of nonlinear
 turbulent fluctuations during an early phase of evolution. The
 initial condition in combination with the random forcing leads to the
 state depicted as figure 1.  In Figure \ref{current}, we follow the
 evolution of both $J_\parallel$ and $J_\perp$ components of the
 currents to quantitatively measure their progressive
 development. Clearly, the two components gradually develop and evolve
 self-consistently in our simulations.  This further leads to $J_\perp
 = J -( {\bf J}\cdot {\bf B}/|{\bf B}| ) {\hat{b}}$ - where
 ${\hat{b}} = {\bf B}/ |{\bf B}|$ is the unit vector along ${\bf B}$,
 - which is non-zero, thus proving conclusively that the resultant
 state is non force-free. For force-free state, $J_{\perp}$ is zero

 The corresponding decay rates associated with turbulent relaxation of
the rugged ideal invariant of MHD, viz, magnetic helicity ($K=\int
{\bf A} \cdot {\bf B} ~d{\bf v}$), magnetic energy ($E= 1/2 \int B^2
~d{\bf v}$) and energy dissipation rate($R=\eta \int J^2 ~d{\bf v}$)
are shown simultaneously in Figure \ref{fig2}. Clearly, the magnetic
energy $E$ decays faster than magnetic helicity $K$, and energy
dissipation ($R$) decays even faster than the two invariants. This
state corresponds to a minimun dissipation in which selective decay
processes lead to the faster decay rates of the magnetic energy (when
compared with the magnetic helicity decay rates).  Furthermore, the
time evolutions of the volume averages of global heliciy, magnetic
energy and the dissipation rates are plotted in Figs 3. Thus, these
quantities are averaged over the fluctuations, and hence they show a
regular behavior.  So the decay of the global heliciy, magnetic energy
and the dissipation rates are not due to any linear decay – they are
the results of nonlinear turbulence in the system.  The selective
decay processes (in addition to dissipations) depend critically on the
cascade properties associated with the rugged MHD invariants that
eventually govern the spectral transfer in the inertial range.  This
can be elucidated as follows (Biscamp, 2003). Magnetic vector
potential in 3D MHD dominates, over the magnetic field fluctuations,
at the smaller Fourier modes, which in turn leads to a domination of
the magnetic helicity invariant over the magnetic energy. On the other
hand, dissipation occurs predominantly at the higher Fourier modes
which give rise to a rapid damping of the energy dissipative quantity
$R$. A heuristic argument for this process can be formulated in the
following way.  The decay rates of helicity $K$ and dissipation rate
$R=
\int_V\eta j^2 dV$ in the dimensionless form, with the magnetic
field Fourier decomposed as ${\bf B}({\bf k},t)= \Sigma_k {\bf
b_k}\exp(i{\bf k}\cdot{\bf r}) $,

\begin{equation}\label{comp1}
\frac{d K}{dt}= -\frac{2\eta}{S}\Sigma_k k {\bf b_k}^2;\qquad
\frac{d R}{dt}= -\frac{2\eta^2}{S^2}\Sigma_k k^4 {\bf b_k}^2
\end{equation}
where $S= \tau_R/\tau_A$, is the Landquist number, $\tau_R,
\tau_A$ are the resistive and Alfv\'{e}n time scales,
respectively. The Landquist number in our simulations varies
between $10^6$ and $10^7$. We find that at scale lengths for which
$k\approx S^{1/2}$, the decay rate of energy dissipation is $\sim
O(1)~1$ . But at these scale lengths, helicity dissipation is only
$\sim O(S^{-1/2})\ll 1$. This physical scenario is further
consistent with our 3D simulations. Interestingly, the decay rates
of kinetic energy of turbulent fluctuations are initially higher
than the magnetic energy. Hence the ratio of magnetic to kinetic
energies shows a sharp rise in the initial evolution, as shown in
Figure \ref{fig3}. However, as the evolution progresses, the two
decay rates become identical and the ratio  approaches eventually
a constant value. Another important outcome to emerge from our
investigations is that a state corresponding to the minimum
dissipation rates, is more plausible in a driven dissipative
plasma. So, we may conclude notwithstanding the Taylor hypothesis
of {\it force-free state}, our simulations clearly demonstrate
that the nonlinear selective decay processes lead the plasma
fluctuations to relax towards a {\it non-force-free state} in a
rather natural and self-consistent manner.

\section{Summary, Conclusion and Future Work}

We have demonstrated through a fully self-consistent 3-D numerical
simulation that  a compressible, dissipative magneto-plasma driven
by large-scale perturbations,  can give rise to a self-organized
state, which is not force-free. This is corroborated by the
presence  of a perpendicular component  of the current $J_\perp$
in the self-organized state. In addition,  the comparative  decay
rates of  global helicity,  magnetic energy and the (Ohmic)
dissipation  rate amply  indicate that dissipation rate can also
serve as an effective minimizer for a driven dissipative  plasma.

It is to be noted that the underlying system has a finite
dissipation. When the rate of dissipation exceeds that of the forcing,
dissipative processes dominate the evolution. In such a case, both the
energy and helicity decreases rapidly. Note that dissipation is
necessary in our simulation to validate the hypothesis of minimum
dissipation rates that lead eventually to a self-organization state in
a dissipative MHD plasma by selectively operating on magnetic energy
and helicity. In the event of initially zero fluctuations, the forcing
leads to populate the turbulent spectra during the early phase of
evolution, which then decay owing to the finite dissipation in the
system. The final results in such cases give rise to the same
conclusions that are described in the context of MDR state elsewhere
in our paper.

Further, we would like to indicate that our results are in line
with Prigogine's idea of the emergence of  new type of
self-organized states called 'Dissipative Structures' for systems
far from equilibrium, where the non-linear interactions of
fluctuations create ordered state in the system. A detailed
investigations on the dissipative structures in plasma will be
undertaken as our future works.

\section{Acknowledgement}

We gratefully acknowledge partial supports from NASA LWS grant
NNX07A073G, and NASA grants NNG04GF83G, NNG05GH38G, NNG05GM62G, a
Cluster University of Delaware subcontract BART372159/SG, and NSF
grants ATM0317509, and ATM0428880.

\newpage

\begin{figure}
\includegraphics[width=14cm]{currents.ps}% Here is how to importEPS art
\caption{\label{current}Evolution of currents associated with the parallel
and perpendicular components of the magnetic field fluctuations in
driven MHD plasma system.  The evolution of a finite $J_\perp$
establishes the fact that the relaxed state is a non-force free state,
i.e., ${\bf J}\times {\bf B} \neq 0 $}
\end{figure}

\begin{figure}
\includegraphics[width=14cm]{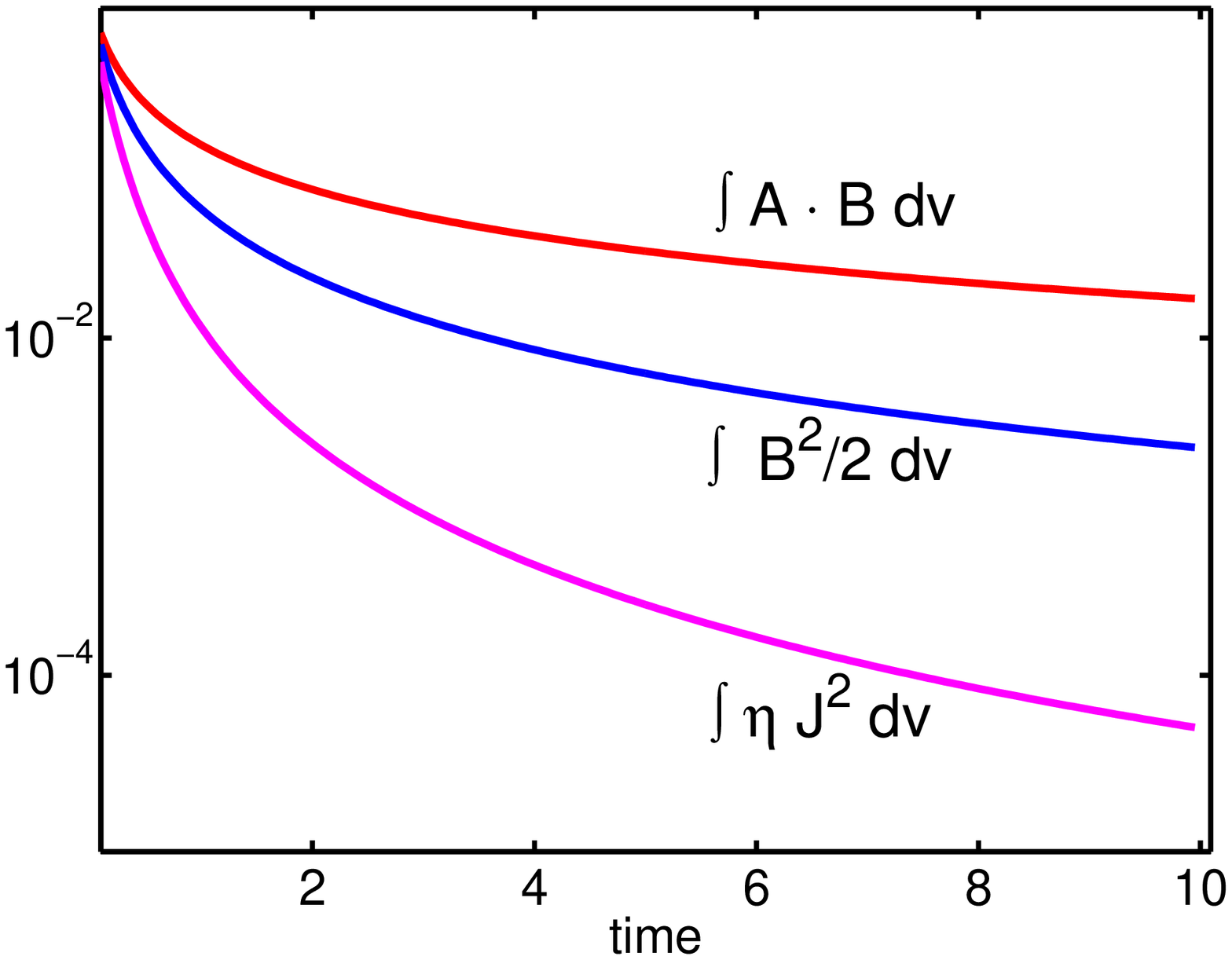}% Here is how to importEPS art
\caption{\label{fig2}Evolution of decay rates associated with
 turbulent relaxation of the rugged ideal invariant of MHD, viz,
 magnetic helicity ($K=\int {\bf A} \cdot {\bf B} ~d{\bf v}$),
 magnetic energy ($E= 1/2 \int B^2 ~d{\bf v}$) and dissipative current
 ($R=\eta \int J^2 ~d{\bf v}$) are shown simultaneously.}
\end{figure}

\begin{figure}
\includegraphics[width=14cm]{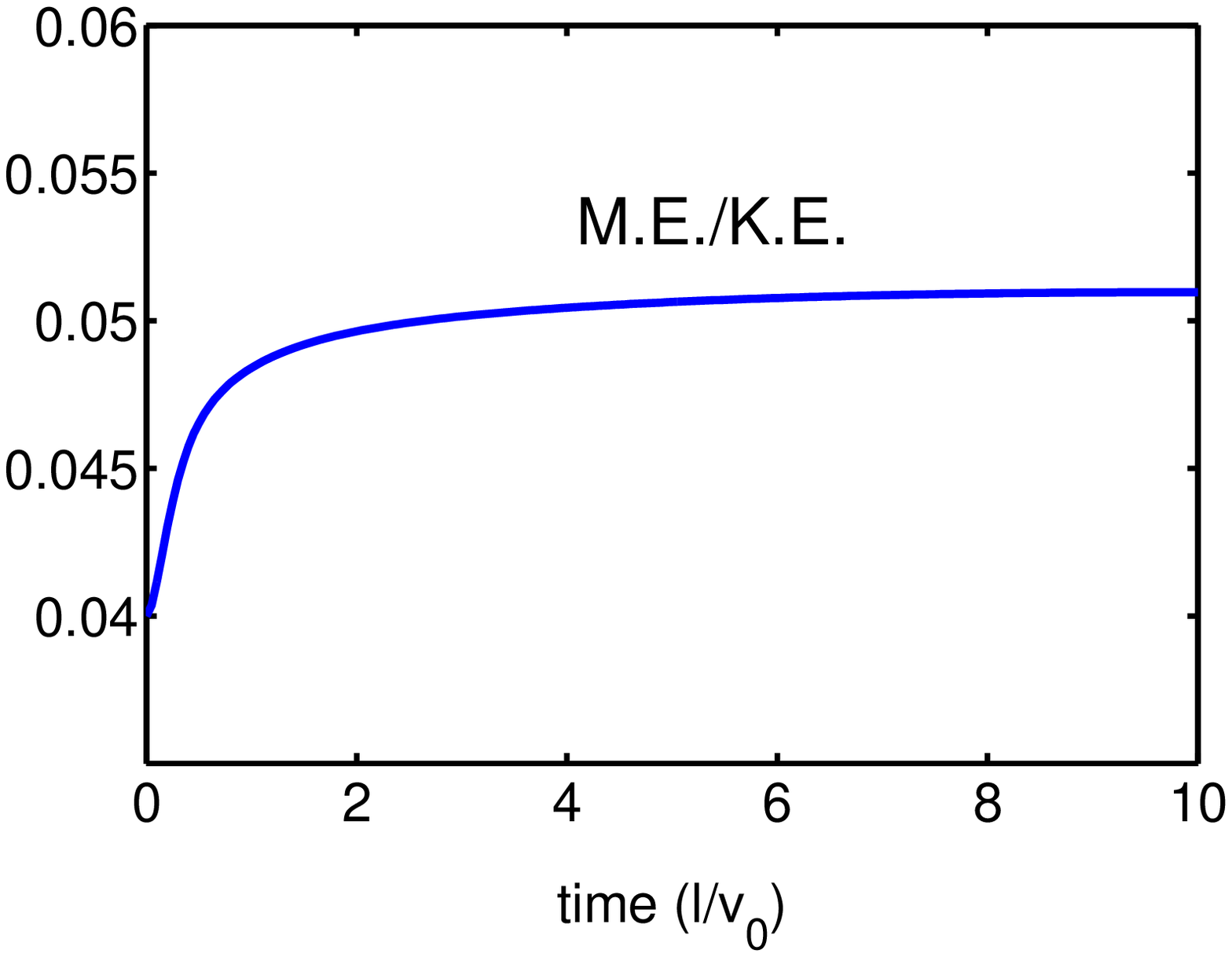}% Here is how to importEPS art
\caption{\label{fig3} The decay rates of kinetic energy of
turbulent fluctuations are initially higher than the magnetic
energy. Hence the ratio of magnetic to kinetic energies shows a
sharp rise in the initial evolution. The two decay rates
eventually become identical thereby leading to a constant value of
the ratio.}
\end{figure}

\label{lastpage}
\end{document}